\documentclass[aps, prx, a4paper, reprint, superscriptaddress, floatfix]{revtex4-2}

\usepackage{amsmath, amssymb}
\usepackage{color, hyperref}
\usepackage[all]{hypcap} 
\usepackage{graphicx}
\usepackage{xcolor}
\usepackage{placeins}
\usepackage[normalem]{ulem}
\usepackage{bm}
\usepackage{siunitx}

\definecolor{MyDarkGreen}{rgb}{0,0.6,0}
\definecolor{MyDarkBlue}{rgb}{0,0,0.8}
\definecolor{MyDarkRed}{rgb}{0.6,0,0.3}
\hypersetup{breaklinks=true,colorlinks=true,plainpages=true,linktocpage=true,linkcolor=MyDarkBlue,citecolor=MyDarkGreen,urlcolor=MyDarkRed,pdfborder={0 0 0},%
pdfauthor={Markus Ilchen},%
pdftitle={Helium paper}%
}

\newlength{\figurewidth}
\begin{document}

\title{
Dichroic Electron Emission Patterns from Oriented Helium Ions
}

\author{Niclas Wieland}
\affiliation{Department of Physics, Universit\"at Hamburg, 22607 Hamburg, Germany}
\email{niclas.wieland@uni-hamburg.de}

\author{Klaus Bartschat}
\affiliation{Department of Physics and Astronomy, Drake University, Des Moines, Iowa 50311, USA}

\author{Filippa Dudda}
\affiliation{Department of Physics, Universit\"at Hamburg, 22607 Hamburg, Germany}

\author{Ren\'e Wagner}
\affiliation{European X-Ray Free-Electron Laser Facility, 22869 Schenefeld, Germany}
\affiliation{Department of Physics, Universit\"at Hamburg, 22607 Hamburg, Germany}

\author{Philipp Schmidt}
\affiliation{European X-Ray Free-Electron Laser Facility, 22869 Schenefeld, Germany}

\author{Carlo Callegari}
\affiliation{Elettra-Sincrotrone Trieste S.C.p.A., 34149 Basovizza, Trieste, Italy}

\author{Alexander Demidovich}
\affiliation{Elettra-Sincrotrone Trieste S.C.p.A., 34149 Basovizza, Trieste, Italy}

\author{Giovanni De Ninno}
\affiliation{Elettra-Sincrotrone Trieste S.C.p.A., 34149 Basovizza, Trieste, Italy}

\author{Michele Di Fraia}
\affiliation{Elettra-Sincrotrone Trieste S.C.p.A., 34149 Basovizza, Trieste, Italy}

\author{Jiri Hofbrucker}
\affiliation{Helmholtz-Institut Jena, Fröbelstieg 3, D-07743 Jena, Germany}

\author{Michele Manfredda}
\affiliation{Elettra-Sincrotrone Trieste S.C.p.A., 34149 Basovizza, Trieste, Italy}

\author{Valerija Music}
\affiliation{European X-Ray Free-Electron Laser Facility, 22869 Schenefeld, Germany}
\affiliation{Institut f{\"u}r Physik und CINSaT, Universit{\"a}t Kassel, 34132 Kassel, Germany}
\affiliation{Deutsches Elektronen-Synchrotron DESY, Notkestr. 85, 22607 Hamburg, Germany}

\author{Oksana Plekan}
\affiliation{Elettra-Sincrotrone Trieste S.C.p.A., 34149 Basovizza, Trieste, Italy}

\author{Kevin C. Prince}
\affiliation{Elettra-Sincrotrone Trieste S.C.p.A., 34149 Basovizza, Trieste, Italy}

\author{Daniel E. Rivas}
\affiliation{European X-Ray Free-Electron Laser Facility, 22869 Schenefeld, Germany}

\author{Marco Zangrando}
\affiliation{Elettra-Sincrotrone Trieste S.C.p.A., 34149 Basovizza, Trieste, Italy}
\affiliation{CNR Istituto Officina dei Materiali, Laboratorio TASC, 34149 Basovizza, Trieste, Italy}

\author{Nicolas Douguet}
\affiliation{Department of Physics, University of Central Florida, Orlando, Florida 32816, USA}

\author{Alexei N. Grum-Grzhimailo}
\affiliation{Skobeltsyn Institute of Nuclear Physics, Lomonosov Moscow State University, 119991 Moscow, Russia}

\author{Michael Meyer}
\affiliation{European X-Ray Free-Electron Laser Facility, 22869 Schenefeld, Germany}

\author{Markus Ilchen}
\affiliation{Department of Physics, Universit\"at Hamburg, 22607 Hamburg, Germany}
\affiliation{Deutsches Elektronen-Synchrotron DESY, Notkestr.~85, 22607 Hamburg, Germany}
\affiliation{European X-Ray Free-Electron Laser Facility, 22869 Schenefeld, Germany}

\begin{abstract}
We report a joint experimental and theoretical study using a 
combination of polarization-controlled free-electron-laser (FEL) and near-infra\-red (NIR) pulses in a synchronized two-color photo\-ionization scheme.
Excited He$^+$ ions, created by extreme ultraviolet (XUV) circularly polarized radiation from the XUV-FEL FERMI in the oriented $3p\, (m\!=\!+1)$ state, are exposed to circularly polarized 784-nm NIR radiation with peak intensities from $10^{12}\,\rm W/cm^2$ to $\rm 10^{13}\,W/cm^2$.
The angular distribution of the ejected electrons exhibit a strong dichroism depending on 
the NIR intensity. 
While the co-rotating case is defined by a single path, 
for the counter-rotating case, 
there are two dominant pathways whose relative strength and phase difference are determined.
\end{abstract}

\maketitle

The transfer of angular momentum from circularly polarized photons to electrons in multi\-photon ionization represents a fundamental quantum process with broad implications for precision measurements, quantum control, and attosecond science~\cite{PhysRevD.99.096017, doi:10.1021/acs.jctc.4c00225, PhysRevLett.120.043202, ilchen2021site, lux2015photoelectron, eckart2018ultrafast}. Recent advances in free-electron-laser (FEL) technology and quantum-sensing applications have intensified interest in understanding and controlling these helicity-dependent interactions, particularly for their potential to enable novel forms of photo\-electron-angular-distribution (PAD) control~\cite{deninnoSingleshotSpectrotemporalCharacterization2015, doi:10.1126/sciadv.add2349, Ilchen17, Mazza2014, hofbrucker2018maximum, ilchen2025opportunities,D1CP05833A,D2CP01009G}. While single-photon circular dichroism is well understood through dipole selection rules, multi\-photon pathways involving intense circularly polarized light may create complex interference patterns between different angular-momentum channels that depend on field intensity and helicity configuration.

Theoretical considerations predict that the partial-wave composition of photo\-electrons should exhibit distinct dependencies on intensity when the multi\-photon ionization involves co-rotating or counter-rotating photon helicities relative to the initial electronic orbital angular momentum. In co-rotating configurations, where photon and electron angular momenta align, the dominant photo\-electron partial wave should remain intensity-independent due to angular-momentum conservation constraints. Conversely, counter-rotating configurations are predicted to exhibit a strong intensity dependence due to changing inter\-ference of different multi\-photon pathways \cite{Ilchen17, Wagner2024}. 

\begin{figure}[!t]
\includegraphics[width=0.47\textwidth]{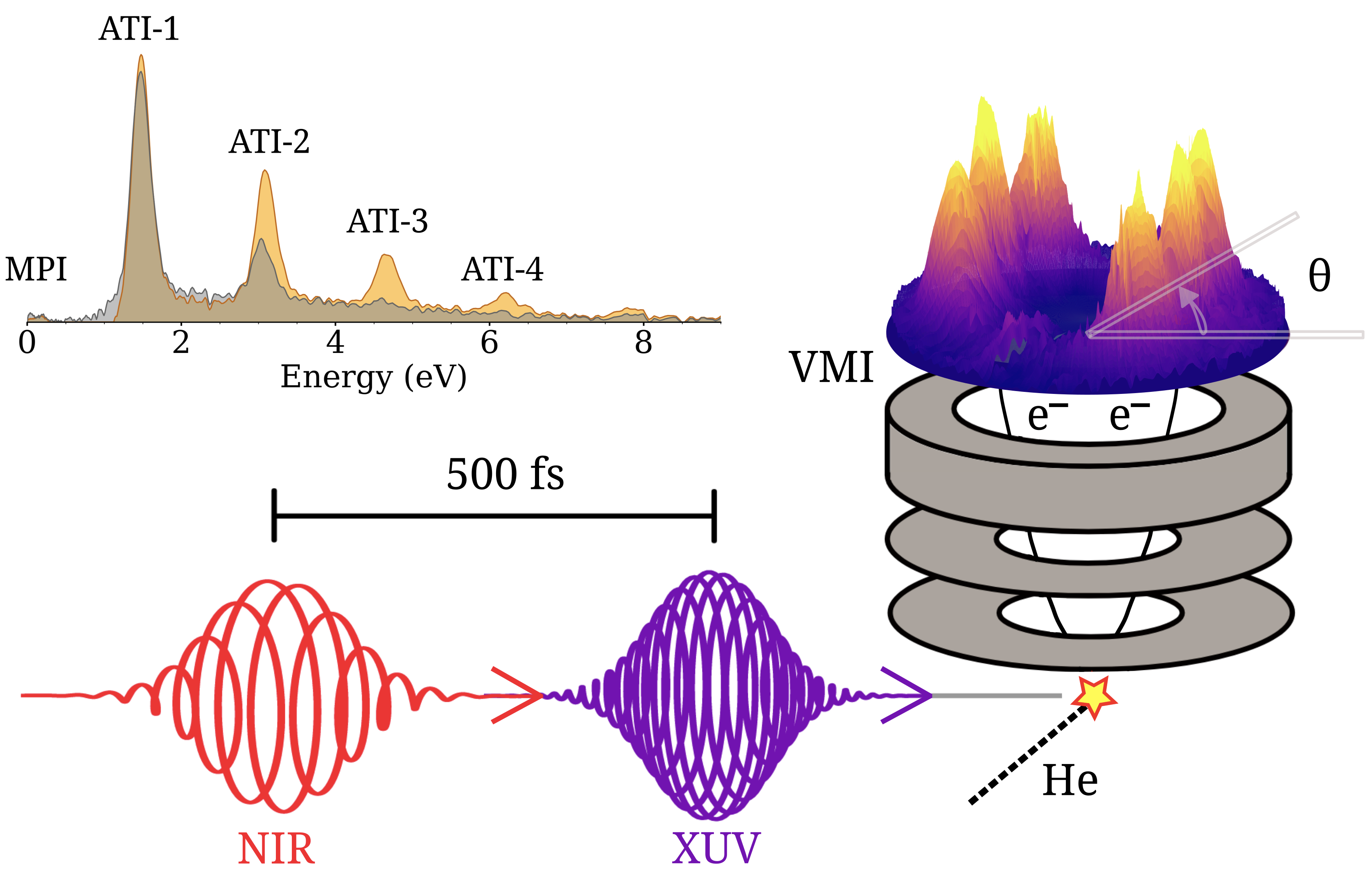}
\label{fig:CD_RadialProfile}
\caption{Illustration of the experimental setup. Circularly polarized XUV pulses 
are tuned to the 
$3p$ resonance of He$^+$, 
which is excited by sequential two-photon absorption within the FEL pulse duration of 75\,fs\,$\pm$ 25\,fs (FWHM). After a delay of about 500~fs, circularly polarized NIR laser light with a mean photon energy of 1.58\,eV (784\,nm) and a bandwidth (FWHM) of 26\,meV (13\,nm) impinges on the sample and produces photo\-electrons via absorption of several NIR photons. For a laser intensity of $\approx 1.1 \times 10^{13}\,\rm W/cm^2$, the  emission spectrum (top left, grey: co-rotating XUV and NIR pulse, yellow: counter-rotating) at the detector location of the VMI shows the ATI-1 to ATI-4 signal for co- and counter-rotating fields.}
\end{figure}

Using a two-color scheme that overlaps an optical laser and an FEL pulse in space and time, we pump a target, helium ions, in a sequential-ionization scheme into the oriented He$^+(3p, m=+1)$ intermediate state. We then provide the missing energy for the double ionization via a multi\-photon process of the optical-laser photons. The minimum number of photons to reach the continuum from this state is four (see multiphoton-ionization peak ``MPI" in Fig.~\ref{fig:CD_RadialProfile}). As shown in the past, AC Stark shifts can dramatically influence the population probability of this pump process and thereby obscure a deeper understanding of the underlying dichroic phenomena \cite{Ilchen17, Wagner2024, grum2019}. To mitigate this influence and concentrate on the dichroic multiphoton processes at hand, we use a temporally separated pulse scheme, where the optical laser pulses are delayed by 500~fs with respect to the FEL pulses, thus decoupling the preparation process of the oriented ionic state  and the subsequent multiphoton processes while maintaining full control over both photon helicity and intensity~\cite{Wagner2024}. 

In order to study the dichroic effects over an extended intensity range, we focus on the first above-threshold ionization (ATI) signal (see Fig.~\ref{fig:CD_RadialProfile}) that requires at least five NIR photons to form. The lowest-order ionization peak of the oriented excited ion moves rapidly below threshold due to ponderomotive shifts already at moderate intensities (see \cite{Wagner2024}). In contrast, the higher order ATIs provide less statistics and exclude the lower intensity range.  This makes the \hbox{ATI-1} peak the ideal candidate to study dichroic partial-wave compositions and phase relations between them. 
We can thereby perform a systematic mapping of PADs across a large intensity range relevant to multi\-photon ionization 
and establish quantitative benchmarks for helicity-dependent quantum control.

In this Letter, 
we present a  
precise measurement of helicity-dependent interfering partial-wave contributions as a function of intensity in multi\-photon ionization, thereby demonstrating detailed mapping of dichroic photo\-electron emission patterns from an oriented He$^+(3p)$ state. Using temporally separated circularly polarized extreme ultraviolet (XUV) and near-infrared (NIR) pulses, we systematically characterize PADs of the first ATI across NIR intensities from $10^{12}\,\rm W/cm^2$ to $ 10^{13}\,\rm W/cm^2$ for both co-rotating and counter-rotating helicity configurations. 
This allows us to determine the relative strength as well as the phase between the two dominating paths for the counter-rotating case, thereby extending the goal of performing a ``complete experiment''~\cite{PAO-Book} from the one-photon to the multi\-photon regime. 

The experiments were performed at the {L}ow-{D}ensity {M}atter (LDM) 
end-station~\cite{Lyamayev_2013} of the seeded free-electron laser
FERMI~\cite{allaria2012}. For a detailed description of the laser parameters, see Wagner {\it et al.}~\cite{Wagner2024}. However, in the present work the peak intensities are shifted upwards by $1.4 \times 10^{12}\,\rm W/cm^2$ with respect to ~\cite{Wagner2024} to account for an improved understanding of the experimental conditions and in light of the theoretical findings.

In the present study, PADs were measured with a velocity-map-imaging (VMI) detector as shown in Fig.~\ref{fig:CD_RadialProfile}. For our geometry, the PAD can be fitted to the form
\begin{equation}\label{eq:angdist}
\frac{d^2\sigma}{dEd\theta} \;(E,\theta,I)  \;\;\;\propto \sum_{n=2,4,6, ...} 1 + \beta_n(E,I) P_{n}(\cos(\theta)).
\end{equation}
Here $E$ is the ejected-electron energy, 
$\theta$ is the detection angle on the VMI, 
$I$ is the NIR peak intensity,  
and $P_{n}(\cos(\theta))$ denotes a standard Legendre polynomial. For $180^\circ \le \theta \le 360^\circ$, we use $360^\circ - \theta$ as the equivalent angle in the above formula.  Experimentally, quadrant mirroring of the raw data was used to enhance the data quality. The inversion of the electron-projection data was performed with the well-established Abel algorithm \cite{vrakking2001iterative}.

Due to the symmetry of our detection scheme and the assumption that no chiral or non-dipole effects play a role under the present conditions, only even-rank asymmetry parameters
$\beta_n$ can contribute to the sum~(\ref{eq:angdist}).
In principle, one would expect $\beta_n$ to depend on both the 
electron energy and the peak laser intensity.  However, we will see below that 
this dependence is not as general as one might expect at first sight.

\begin{figure*}[!t]
\includegraphics[width=0.85\textwidth]{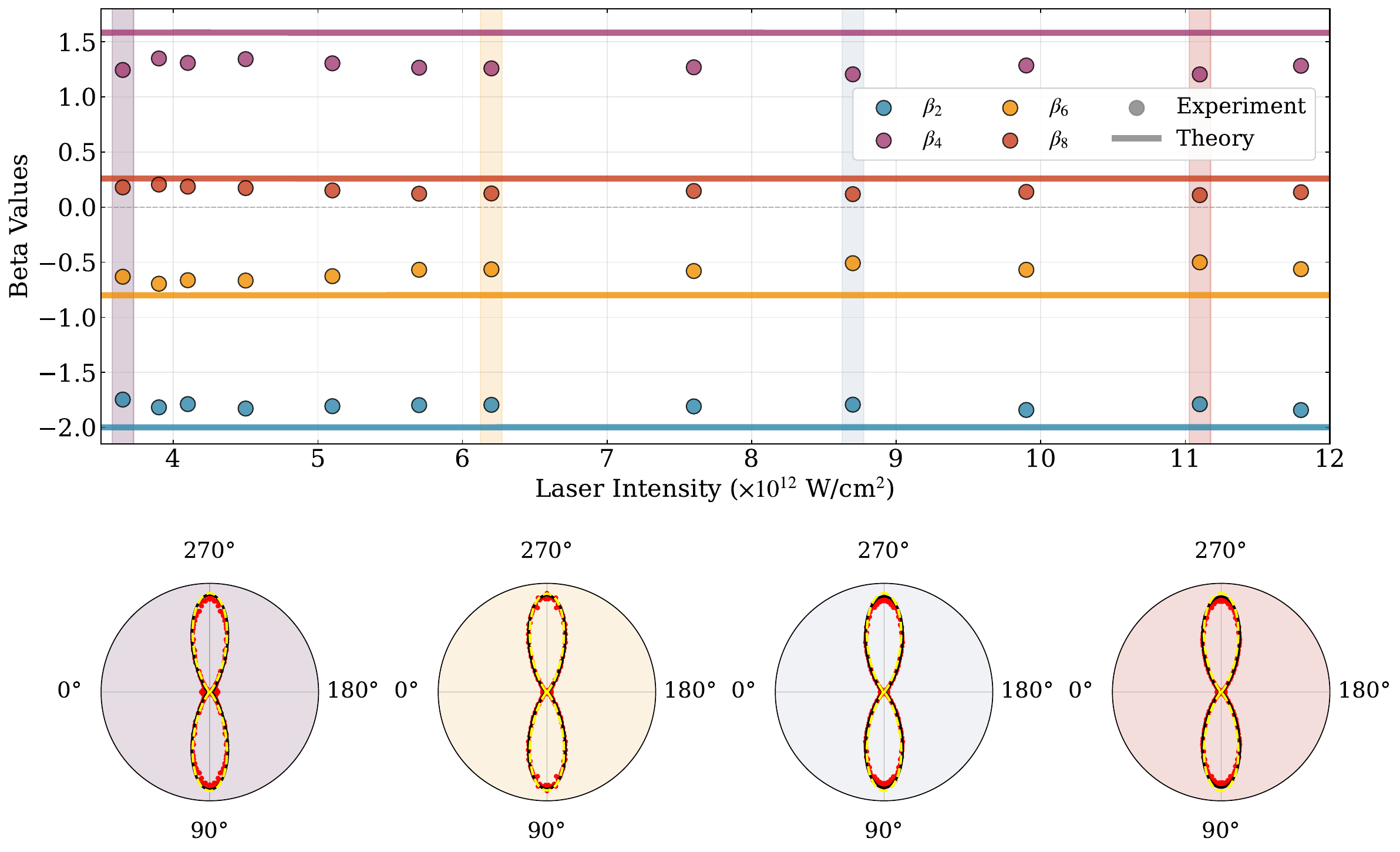}
\label{fig:beta_eva_theory_experiment-co}
\caption{Top: Beta parameters ($\beta_2$, $\beta_4$, $\beta_6$, $\beta_8$) as a function of laser intensity from $3.5$ to $12 \times 10^{12}\,\rm W/cm^2$ for the ATI-1 peak in the co-rotating case. Experimental data (filled circles) obtained by averaging over all quadrant combinations are compared with theoretical predictions (lines). Colored vertical boxes indicate the intensities selected for a detailed polar-plot analysis. Bottom: Polar plots showing PADs at four representative intensities. The red points represent the experimental data, the black solid lines show experimental fits, and the yellow dashed lines are the relative theoretical predictions. The background colors correspond to the intensity markers in the top panel.}
\end{figure*}

The calculations were performed with the same method and computer code described in detail by Wagner {\it et al.}~\cite{Wagner2024}.
Briefly, we solved the time-dependent Schr\"odinger equation (TDSE) for the active electron starting in the
$3p\, (m\!=\!+1)$ state of He$^+$. 
Due to the 500-fs delay between the generation of the excited He$^+$ ion, sequential
ionization of the He$(1s^2)$ ground state and excitation of the remaining electron from the $1s$ to the $3p$ orbital, and the multi-photon ionization of the excited state by the intense NIR laser,
this is a true one-electron problem. The non\-relativistic 
orbitals are known analytically, and the continuum states to project the final-state
wave function
on to obtain the ionization signal are pure Coulomb functions.

Figure~\ref{fig:beta_eva_theory_experiment-co} shows the parameters~$\beta_n$ for $n=2,4,6,8$ 
obtained for the co-rotating case, i.e., the same helicity of the XUV and NIR laser beams.
Without loss of generality, we assume the helicity of the XUV photon to
be $+1$, i.e., we prepare the He$^+(3p,m=+1)$ state.  For the co-rotating case,
the dominant pathway after the absorption of $N$ NIR photons, therefore, is to
reach a state whose PAD is proportional to  
\begin{equation}
|Y_{N+1}^{N+1}(\theta,\phi=0^\circ,180^\circ)|^2 \propto |\sin(\theta)|^{2N+2}.
\end{equation}

The ATI-1 line is therefore characterized by a pathway resulting in a continuum electron with angular momentum $\ell = 6$, since 5 NIR photons are required for the ionization process.
Note that this result is entirely \emph{independent} of the NIR intensity.
Consequently, the $\beta_n$ parameters cannot be independent of each other, but are the unique expansion coefficients of 
$|\sin(\theta)|^{2N+2}$ into a series of (orthogonal) Legendre polynomials.

\begin{figure*}[!t]
\includegraphics[width=0.85\textwidth]{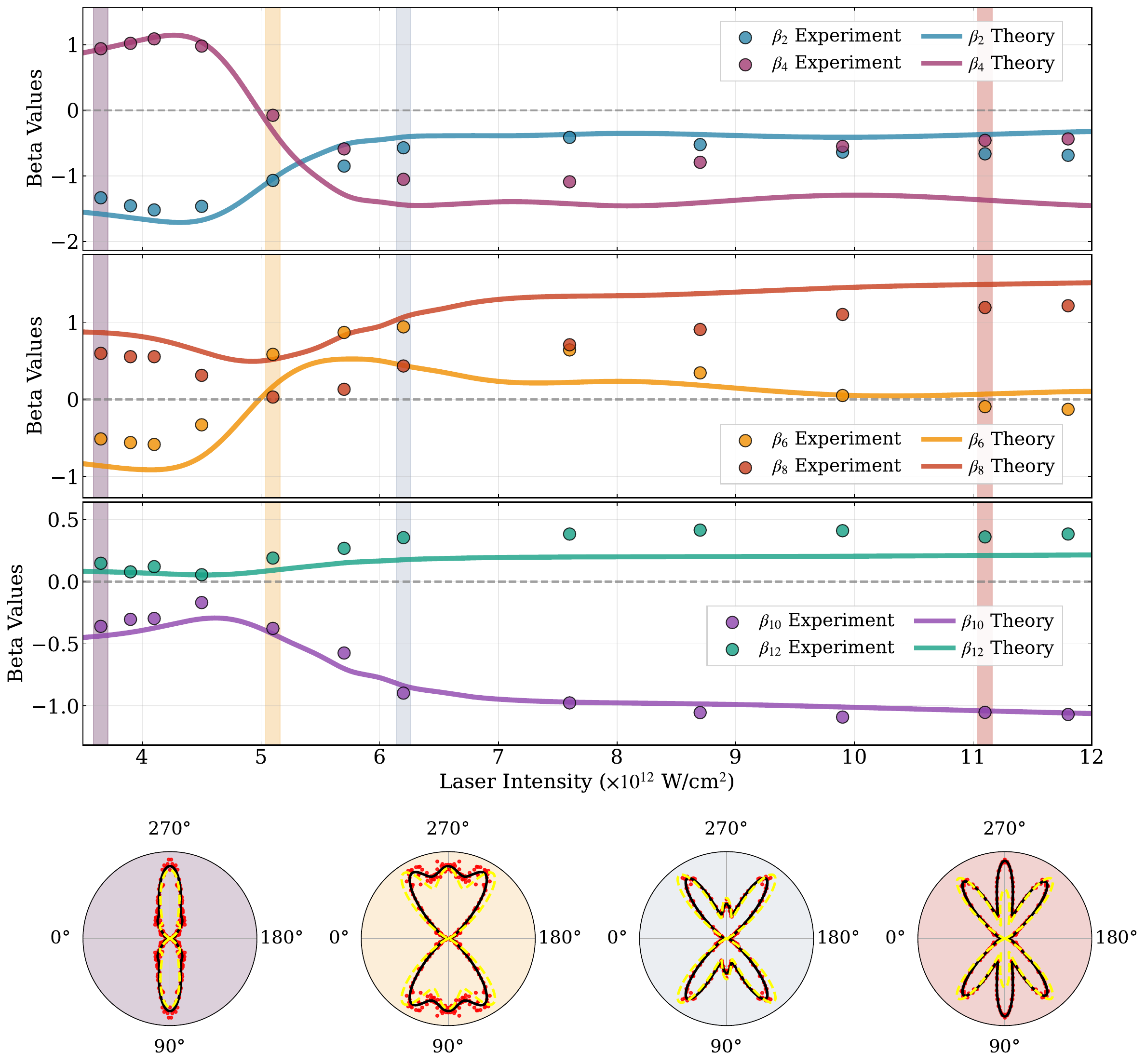}
\label{fig:beta_eva_theory_experiment-counter}
\caption{Same as Fig.~\ref{fig:beta_eva_theory_experiment-co} (with expansions up to $\beta_{12}$ added) for the counter-rotating case. The selected angular distributions reveal a dependence on both the relative strength and the phase between the two dominant interfering pathways. The areas around \num{0}\si{\degree} and \num{180}\si{\degree} are affected by beam-induced artifacts which requires the exclusion of $\pm$\num{15}\si{\degree}. 
}
\end{figure*} 

This result is, indeed, verified by both the experimental data, albeit with
some minor fluctuations, and the theoretical
predictions shown in Fig.~\ref{fig:beta_eva_theory_experiment-co}.  Both data sets are characterized by a constant value for the $\beta_n$ parameters.
We only show $\beta_2$, $\beta_4$, $\beta_6$, and $\beta_8$ for this case,
even though higher terms appear but are too small to be experimentally
distinguishable from zero.

Moving on to the counter-rotating case, we see an entirely different pattern in Fig.~\ref{fig:beta_eva_theory_experiment-counter}. For this case, we experimentally determined $\beta_2$, $\beta_4$, $\beta_6$, $\beta_8$, $\beta_{10}$, and $\beta_{12}$ as clearly being non\-zero in general.  Also, there is a strong intensity dependence for all of them, resulting in a drastic change of the PAD from the lowest intensity of $3.5 \times 10^{12}\,\rm W/cm^2$ to the next selected one ($5.1 \times 10^{12}\,\rm W/cm^2$) and, finally, to intensities above 
$6 \times 10^{12}\,\rm W/cm^2$. Such a strong intensity dependence of the PAD, accompanied by a rapid change over a small range of NIR intensity, was unexpected in light of the fact that i)~the angle-integrated dichroism (cf.\ Fig.~9 of~\cite{Wagner2024}) is smoothly decreasing in the corresponding intensity range from about 0.4 at lowest intensity to about zero at high peak intensities and ii)~the pattern for the co-rotating case does not change at all.

The fact that, in contrast to the co-rotating case, there is an intensity dependence for the counter-rotating case can be understood from the two pathways that should predominantly contribute to the PAD, namely pathways with angular momentum $\ell = 6$ and $\ell = 4$ for the outgoing electron \cite{Ilchen17}.    
These partial waves will interfere to form a state that can be represented by the linear combination 
\begin{equation} \label{eq:counter-state}|A| Y_6^{-4} + |B|\exp{({\rm i}\delta}) \,Y_4^{-4}.
\end{equation}  
Hence, the measured angular distribution will contain the absolute strengths 
$|A|^2$ and $|B|^2$ that determine the generalized multi\-photon  cross section, as well as the interference term $|A|\,|B|\,\cos\delta$.
The $\beta_n$ parameters then become known functions of the relative strength $|B|/|A|$ and the phase~$\delta$. 
Considering this, and the excellent agreement between the measured and calculated PADS (cf. Figs.~\ref{fig:beta_eva_theory_experiment-co} and~\ref{fig:beta_eva_theory_experiment-counter}), we refrain from further analyzing the $\beta_n$ and focus on the phase~$\delta$. 

\begin{figure}[!h]
\includegraphics[width=0.93\columnwidth]{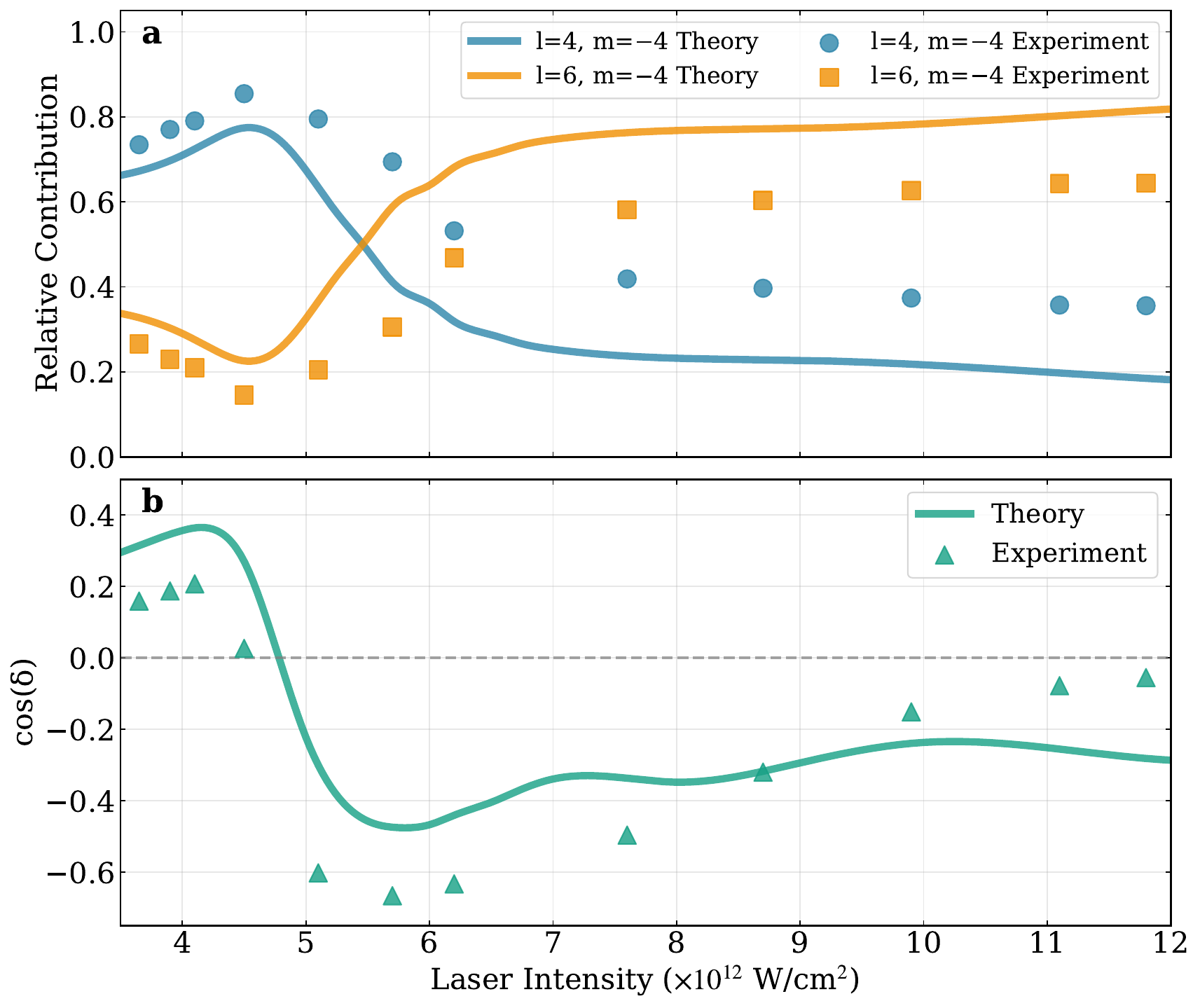}
\label{fig:partial}
\caption{a) Dominant partial-wave contributions~(a) and $\cos(\delta)$~(b) as a function of laser intensity 
for the \hbox{ATI-1} peak in the counter-rotating case.}
\end{figure} 

It is, indeed, possible to determine both the relative strength and the cosine of the phase between the two dominant pathways.
Figure~\ref{fig:partial}(a) shows the experimental data and the theoretical predictions for the relative strengths, normalized to their sum being~1.  We see a strong intensity dependence up to intensities of about $6 \times 10^{12}\,\rm W/cm^2$ with changes of the relative contributions by up to a factor of~4. Above this region of intensity, the contributions are nearly constant, with $\ell = 6$ dominating over $\ell = 4$.  This is expected from the propensity rule of likely adding angular momentum to the ejected electron when its energy is increased~\cite{Fano-propensity}. Figure~\ref{fig:partial}(b) shows $\cos(\delta)$, which  also varies strongly between $+0.4$ and $-0.4$ in the intensity region up to $6 \times 10^{12}\,\rm W/cm^2$ and then remains negative for higher intensities. 

We interpret the observed patterns to originate from the crossing of the relative partial-wave strength and to the strong NIR intensity dependence of the relative phase.  
They \emph{may}, at least to some extent, have to do with the fact that the MPI line moves below threshold with increasing intensity. Hence, it is more likely that the five-photon ATI-1 line is affected by the fourth photon exciting members of the dense Rydberg manifold. 
However, a more quantitative explanation is desirable, and the search for it will be the subject of future studies.

To summarize: We analyzed in detail the dichroic, angle-dependent electron-emission pattern observed after irradiating oriented He$^+(3p)$ ions, produced by circularly polarized FEL laser pulses, with intense NIR light having the same or opposite helicity as the FEL.  As expected, the PAD for the co-rotating cases can effectively be described by a single partial wave for all intensities studied.  Consequently, the intensity-independent $\beta_n$ parameters are determined by addition theorems for the Legendre polynomials.

For the counter-rotating case, on the other hand, interference of two dominating partial-wave contributions leads to more complex PADs.  Interestingly, we found a rapid variation in both the relative strength and the phase as a function of the NIR intensity.  Their determination brought us one step closer to a ``complete experiment'' in a multi\-photon scenario.
Realistically, however, determining the absolute values of the coefficients ~$A$ and~$B$ in Eq.~(\ref{eq:counter-state}), i.e., the absolute generalized photo\-ionization cross section, is currently out of reach. 

In light of the experimental challenges, the agreement between our theoretical predictions and our experimental data is satisfactory. We suspect that less than 100\% circular polarization of the two beams and slight mis\-alignments are mainly responsible for the remaining discrepancies.  We emphasize that focal-point averaging, even if it were necessary, would not resolve the differences in the co-rotating case, since constant $\beta_n$-values are expected independent of the laser intensity. 


In conclusion: Controlling not only the angle-integrated CD but also the angular-resolved PAD via the mutual helicity of the two laser beams and the NIR intensity is expected to be a \emph{general} tool that could also be applied to much more complex systems than the proof-of-principle case presented in this Letter. 

\smallskip


We thank Thomas Pfeifer and Christian Ott (Max Planck Institute for Nuclear Physics, Heidelberg) for fruitful discussions and the technical and scientific staff at FERMI (Elettra-Sincrotrone Trieste) for their support. R.W.\ and M.M.\ acknowledge funding by the Deutsche Forschungs\-gemeinschaft (DFG) in SFB-925, ``Light induced dynamics and control of correlated quantum systems'', project No.\ 170620586. M.I., V.M., and Ph.S.\ received a Peter-Paul-Ewald Fellowship from the Volks\-wagen Foundation. M.I.\ also acknowledges DFG support under project No.\ 328961117 in SFB-1319, ``Extreme light for sensing and driving molecular chirality''. M.I., N.W., and M.M.\ were supported by the DFG Cluster of Excellence ``CUI: Advanced Imaging of Matter'', project No.\ 390715994. Theoretical work was funded by the NSF (grant Nos.\ \hbox{PHY-2408484} [K.B.], \hbox{PHY-2012078} [N.D.]) and the ACCESS supercomputer allocation \hbox{No.~PHY-090031}. Work by A.N.~Grum-Grzhimailo was conducted prior to the war in Ukraine.

\bibliographystyle{apsrev4-2}
\bibliography{He_2022}

\end{document}